\documentclass[fleqn,10pt]{wlscirep}
\usepackage[utf8]{inputenc}
\usepackage[T1]{fontenc}
\usepackage{lineno}
\usepackage{makecell}
\usepackage{multirow}

\usepackage{tabularx, booktabs}
\usepackage{rotating}
\usepackage{array}

\newcommand{\beginsupplement}{%
        \setcounter{table}{0}
        \renewcommand{\thetable}{S\arabic{table}}%
        \setcounter{figure}{0}
        \renewcommand{\thefigure}{S\arabic{figure}}%
        \renewcommand{\figurename}{Supplementary Figure}
        \renewcommand{\tablename}{Supplementary Table}
     }

\title{ProNet DB: A proteome-wise database for protein surface property representations and RNA-binding profiles}

\author[1,$\dag$]{Junkang Wei}
\author[1,$\dag$]{Jin Xiao}
\author[2,3,$\dag$]{Siyuan Chen}
\author[1,]{Licheng Zong}
\author[2,3,*]{Xin Gao}
\author[1,4,5,6,7,*]{Yu Li}
\affil[1]{Department of Computer Science and Engineering (CSE), The Chinese University of Hong Kong (CUHK), 999077, Hong Kong SAR, China}
\affil[2]{Computer Science Program, Computer, Electrical and Mathematical Sciences and Engineering (CEMSE) Division, King Abdullah University of Science and Technology (KAUST), Thuwal 23955-6900, Kingdom of Saudi Arabia}
\affil[3]{KAUST Computational Bioscience Research Center (CBRC), King Abdullah University of Science and Technology, Thuwal 23955-6900, Kingdom of Saudi Arabia}
\affil[4]{The CUHK Shenzhen Research Institute, Shenzhen 518057, China}
\affil[5]{Institute for Medical Engineering and Science, Massachusetts Institute of Technology, Cambridge, MA, USA}
\affil[6]{Wyss Institute for Biologically Inspired Engineering, Harvard University, Boston, MA, USA}
\affil[7]{Broad Institute of MIT and Harvard, Cambridge, MA, USA}

\affil[*]{Corresponding author(s): Xin Gao, xin.gao@kaust.edu.sa; Yu Li, liyu@cse.cuhk.edu.hk}

\affil[$\dag$]{These authors contributed equally to this work}

\begin{abstract}
The rapid growth in the number of experimental and predicted protein structures and more complicated protein structures challenge users in computational biology for utilizing the structural information and protein surface property representation. Recently, AlphaFold2 released the comprehensive proteome of various species, and protein surface property representation plays a crucial role in protein-molecule interaction prediction such as protein-protein interaction, protein-nucleic acid interaction, and protein-compound interaction. Here, we proposed the first comprehensive database, namely ProNet DB, which incorporates multiple protein surface representations and RNA-binding landscape for more than 326,175 protein structures covering 16 model organism proteomes from AlphaFold Protein Structure Database (AlphaFold DB) and experimentally validated protein structures deposited in Protein Data Bank (PDB). For each protein, we provided the original protein structure, surface property representation including hydrophobicity, charge distribution, hydrogen bond, interacting face, and RNA-binding landscape such as RNA binding sites and RNA binding preference. To interpret protein surface property representation and RNA binding landscape intuitively, we also integrate Mol* and Online 3D Viewer to visualize the representation on the protein surface. The pre-computed features are available for the users instantaneously and boost computational biology development including molecular mechanism exploration, geometry-based drug discovery and novel therapeutics development. The server is now available on \textcolor{blue}{\href{https://proj.cse.cuhk.edu.hk/aihlab/pronet/}{https://proj.cse.cuhk.edu.hk/aihlab/pronet/}}.
\end{abstract}
\begin{document}

\flushbottom
\maketitle

\thispagestyle{empty}

\begin{figure*}[t]
\includegraphics[width=0.95\textwidth]{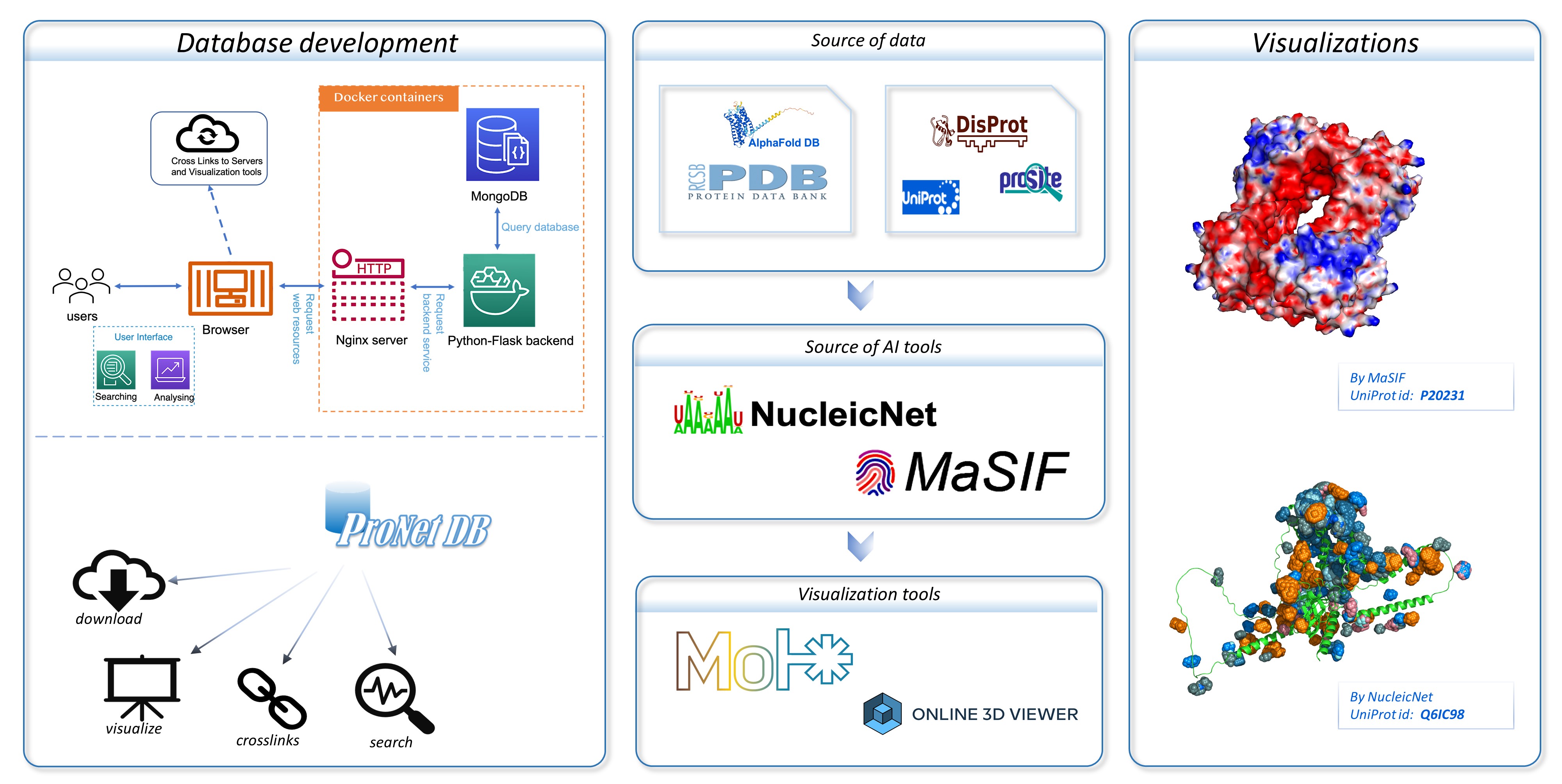}
\centering
\caption{An overview of the ProNet DB and the illustration for two main outputs. The right panel shows the example of the protein surface physicochemical property and RNA binding profiles.}
\label{Structure}
\end{figure*}

\section*{Background \& Summary}

Proteins perform vital functions in a variety of cellular activities, and protein-molecule interactions decipher the complexity of organisms such as gene expression regulation \cite{weirauch2013evaluation}, signal transduction \cite{alipanahi2015predicting} and drug therapy \cite{lu2020recent}. However, the dedicated mechanism of most protein-molecule interactions has not been well illustrated and hinders the development of mechanism exploration and drug discovery. During the process of protein interacting with molecules, molecules are intended to recognize the surface of the protein, such as the hydrophobicity, charge distribution, hydrogen/electron donor, and binding steric hindrance. Thus, a comprehensive and efficient representation of the protein surface is essential to elucidate the mechanism of protein-molecule interaction. For example, Rudden et al. \cite{rudden2019protein} utilize a single volumetric descriptor representing protein surface including electrostatics and local dynamics for protein docking and achieved an average success rate of 54\%. Experimental assessments such as NMR-based measurement \cite{almeida2021protein},  hydrophobic interaction chromatography (HIC) \cite{lienqueo2007current} of protein surface property are time-consuming and costly. Besides, with the presence of AlphaFold2 Protein Structure Database \cite{varadi2022alphafold}, a number of protein structures are determined by computational prediction, indicating that the traditional approaches are unable to handle this series of protein surface property evaluation. 

To overcome the limitations of experimental approaches, several \textit{in silico} methods of protein surface property have been proposed, such as MaSIF \cite{gainza2020deciphering}, FEATURE \cite{halperin2008feature}, and AutoDock \cite{huey2012using}. For example, AutoDock \cite{huey2012using} calculates the atom-wise biochemical property, and FEATURE \cite{halperin2008feature} employs a series of centric shells to represent atoms of the protein with 7.5Å of a grid point with 80 physicochemical properties. MaSIF \cite{gainza2020deciphering} presents a method to encode geometric features (shape index and distance-dependent curvature) and chemical features (hydropathy, continuum electrostatics, and free electrons/protons) on the surface with the geodesic radius of 9Å or 12Å. Despite the availability of those tools for downstream applications, they are not ready-to-use, with the complex running environment and long running time. Also, it is inefficient for each user to run them locally for the same protein, which leads to repetitive work. Theoretically, for the fixed protein structure, the surface representation of the same tool should be the same. Considering that, we build up the database, running MaSIF to encode protein surface physicochemical properties including hydrophobicity, charge distribution, hydrogen bond, interacting face for the protein structure from the experimentally validated database (PDB) and \textit{in silico} database (AlphaFold DB), so that the user can directly use such features for their downstream applications. The successful \textit{de novo} design of protein with learned surface fingerprints revealed that surface property plays a crucial role in function-oriented protein design, and lay the foundation for the development of synthetic biology \cite{gainza2023novo}. 

Similar to the physicochemical property, RNA binding landscape is also an important part of surface property. Direct recognition of RNA motifs on RNA-binding proteins (RBPs) can provide information of protein-nucleic acid interaction \cite{wei2022protein}. For example, the Pumilio/FBF (PUF) family can govern translations by direct base-protein recognition, such as UGUR motifs on RNA transcripts \cite{quenault2011puf}. Thus, the RNA binding profiles of RBPs are the important part to illustrate protein-molecule interaction. In this study, we employed the state-of-the-art deep-learning framework NucleicNet \cite{li2019deep} to predict the binding preference of RNA constituents and the binding sites on protein surface to provide RNA-binding landscape of the protein structure from the experimentally validated database (PDB) and \textit{in silico} database (AlphaFold DB). Although the dataset is based on prediction, we are the first to provide such a ready-to-use database for downstream applications, such as CRISPR-Cas system optimization \cite{tycko2016methods}, RBP-targeting therapeutics discovery \cite{gebauer2021rna}, and aptamer-guided drug delivery system development \cite{alshaer2018aptamer}.

In summary, we proposed a comprehensive database for protein surface feature, ProNet DB, which contains protein surface physicochemical representations and RNA-binding landscape for more than 326,175 protein structures covering the 16 model organism proteomes from AlphaFold DB and PDB. For each protein, we provided the original protein structure, surface property representation including hydrophobicity, charge distribution, hydrogen bond, interacting face, and RNA-binding landscape such as RNA binding site and RNA binding preference. To interpret protein surface property representation and RNA binding landscape intuitively, we also integrate Mol* and Online 3D Viewer to visualize representation on the protein surface. The server now can be assessed at \textcolor{blue}{\href{https://proj.cse.cuhk.edu.hk/aihlab/pronet/}{https://proj.cse.cuhk.edu.hk/aihlab/pronet/}} and future releases will expand the species and property coverage.

\begin{figure*}[t]
\includegraphics[width=0.9\textwidth]{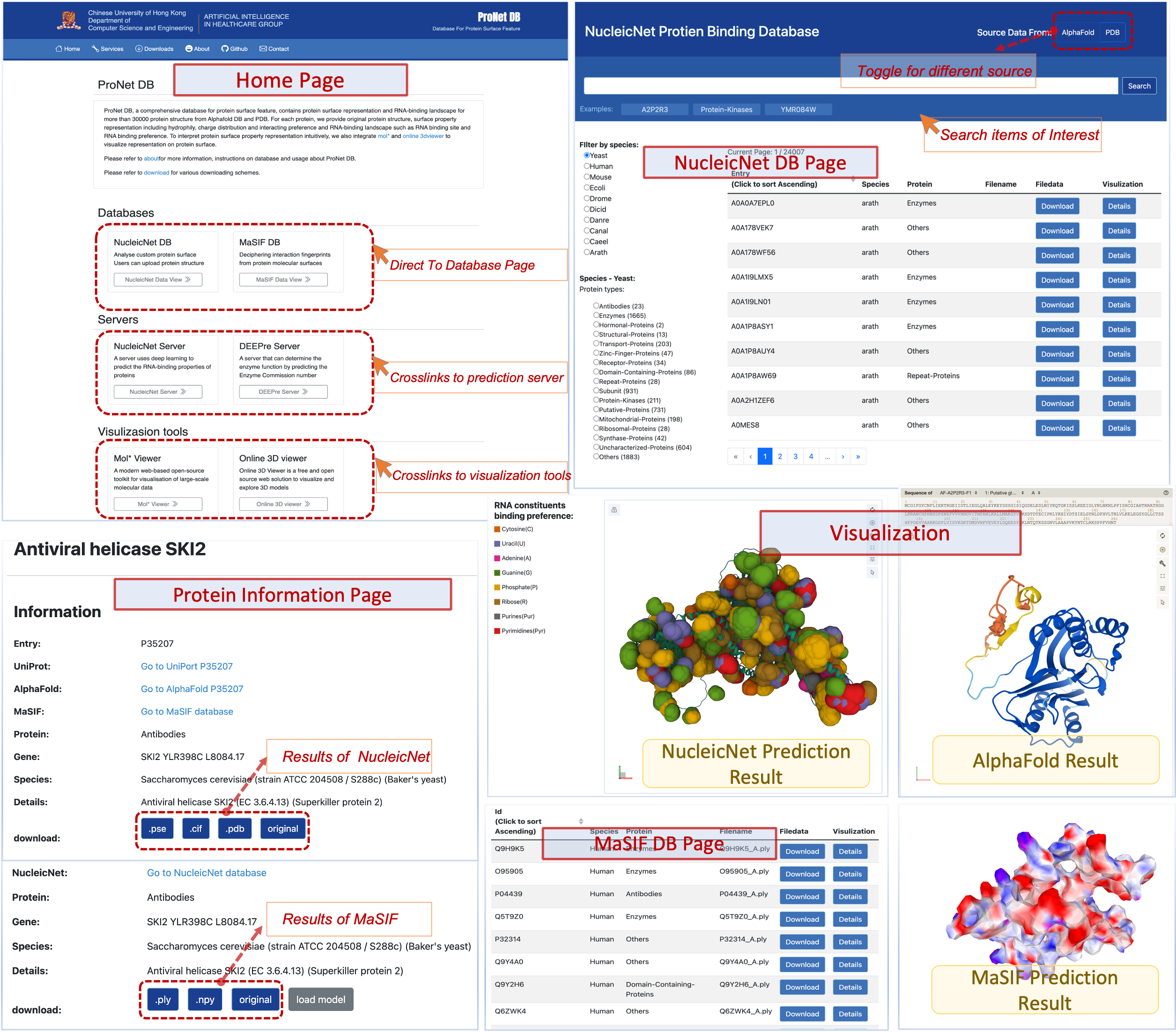}
\centering
\caption{User interface of ProNet DB. \textbf{Top-left}: Home page contains three subsections: servers, databases, and visualization tools. \textbf{Top-right}: NucleicNet DB page. Users can search, filter and view the searched results. On the top-right corner of NucleicNet DB page, a toggle button provides different protein sources. \textbf{Bottom}: Protein information page and visualization details for each item.}
\label{overview_withlabel}
\end{figure*}

\section*{Methods}

\subsection*{Data Source}

We first collected 23,391 protein structures on \textit{Homo sapiens} proteome and 6,042 protein structures on \textit{Saccharomyces cerevisiae} proteome from AlphaFold DB \cite{varadi2022alphafold}. If the corresponding experimentally validated protein structures exist in PDB, we collected the protein structure with the highest resolution from PDB (\textit{Homo sapiens}: 6,030, \textit{Saccharomyces cerevisiae}: 1,160) \cite{burley2017protein}. For further comprehensive database construction, we collected the other 14 model organism proteomes from AlphaFold DB, including \textit{Arabidopsis thaliana}, \textit{Caenorhabditis elegans}, \textit{Candida albicans}, \textit{Danio rerio}, \textit{Dictyostelium discoideum}, \textit{Drosophila melanogaster}, \textit{Escherichia coli}, \textit{Glycine max}, \textit{Methanocaldococcus jannaschii}, \textit{Mus musculus}, \textit{Oryza sativa}, \textit{Rattus norvegicus}, \textit{Schizosaccharomyces pombe} and \textit{Zea mays}. Finally, the proteomes of these model organisms sufficiently expanded ProNet DB protein structure coverage from 33,000 to 333,365 (Table \ref{proteomes}) and led to a more comprehensive and user-friendly database. 

\begin{table}[ht]
\centering
\caption{The model organism proteomes in ProNet DB.}
\begin{tabular}{cccccc}
\hline
ID & Species                                & Name   & Reference Proteome & AlphaFold DB & PDB   \\ \hline
1  & \textit{Arabidopsis thaliana}                   & Arabidopsis   & UP000006548        & 27,434       & -     \\
2  & \textit{Caenorhabditis elegans}        & Nematode worm & UP000001940        & 19,694       & -     \\
3  & \textit{Candida albicans}              & C. albicans   & UP000000559        & 5,974        & -     \\
4  & \textit{Danio rerio}                   & Zebrafish     & UP000000437        & 24,664       & -     \\
5  & \textit{Dictyostelium discoideum}      & Dictyostelium & UP000002195        & 12,622       & -     \\
6  & \textit{Drosophila melanogaster}       & Fruit fly     & UP000000803        & 13,458       & -     \\
7  & \textit{Escherichia coli}              & E. coli       & UP000000625        & 4,363        & -     \\
8  & \textit{Glycine max}                   & Soybean       & UP000008827        & 55,799       & -     \\
9  & \textit{Homo sapiens}                  & Human         & UP000005640        & 23,391       & 6,030 \\
10 & \textit{Methanocaldococcus jannaschii} & M. jannaschii & UP000000805        & 1,773        & -     \\
11 & \textit{Mus musculus}                  & Mouse         & UP000000589        & 21,615       & -     \\
12 & \textit{Oryza sativa}                  & Asian rice    & UP000059680        & 43,649       & -     \\
13 & \textit{Rattus norvegicus}             & Rat           & UP000002494        & 21,272       & -     \\
14 & \textit{Saccharomyces cerevisiae}      & Budding yeast & UP000002311        & 6,040        & 1,160 \\
15 & \textit{Schizosaccharomyces pombe}     & Fission yeast & UP000002485        & 5,128        & -     \\
16 & \textit{Zea mays}                      & Maize         & UP000007305        & 39,299       & -     \\
   &                                        &               &                    & 326,175      & 7190     \\ \hline
\end{tabular}
\label{proteomes}
\end{table}

\subsection*{Protein Surface Physicochemical Property}
MaSIF is a general framework to encode protein surface fingerprints \cite{gainza2020deciphering}. For each protein, it will generate a discretized molecular surface by assigning calculated physicochemical features on every vertex of the surface. In this way, the properties of the protein surface can be clearly represented. As shown in Figure \ref{Structure}, the user can determine which part of the surface area is hydrophilic or hydrophobic, and which part is more likely to interact with other molecules (interacting face). We computed the surface properties by the MaSIF tool for the proteins in our database so that users can obtain the physicochemical property profile for every protein efficiently. These computed features can benefit many downstream tasks a lot including binding site prediction \cite{miotto2021molecular}, protein-protein interaction prediction \cite{gaudelet2021utilizing} and protein design \cite{gao2020deep}. The recent study of protein design based on surface fingerprints confirmed that surface property plays a crucial role in function-oriented protein design, and indicated that such geometric features on protein surface boost protein-centric issue development \cite{gainza2023novo}. 

\subsection*{Protein RNA-binding Profiles}
As RNA-protein interaction is involved in multiple cellular activities, the interaction between RNAs and RBPs plays an important role in understanding cellular activities. The systematic mappings of the RNA-protein interaction for multiple RNA constituents were constructed in ProNet DB. Following the deep-learning framework proposed by NucleicNet \cite{li2019deep}, we acquired the binding preference as well as the binding sites for multiple bases for protein structures in Alphafold DB and PDB, such as Ribose (R), Phosphate (P), Adenine (A), Guanine(G), Cytosine (C), and Uracil (U). The protein RNA binding profiles are further classified into multiple sub-classes for each species based on their diverse protein functions. In ProNet DB, users are able to directly address the protein properties such as RNA backbone composition and binding preference of different bases, which intuitively illustrated the protein RNA-binding landscape and partially revealed the protein surface property. 

\begin{table*}[ht!]
\centering
\caption{An example entry in ProNet DB shows the data content organization of one protein \textit{17-beta-hydroxysteroid dehydrogenase type 1}. An entry has three profiles: \textbf{Basic Profile} contains basic information like the protein names, protein types, gene names, as well as the mapping id to other databases; \textbf{MaSIF Profile} includes the physicochemical properties computed by MaSIF, describing the protein surface features; \textbf{NucleicNet Profile} contains the RNA-binding preference information.}

\begin{tabular}{lll}
\toprule
\multicolumn{2}{l}{Description}  & Example                       \\ 
\midrule
\multirow{11}{*}{Basic Profile}     
& Entry ID  &   P14061                                           \\
& Protein Name  & 17-beta-hydroxysteroid dehydrogenase type 1    \\
& PDB ID      & 1A27                                             \\
& Uniprot ID & P14061                                           \\
& Sequence Length   & 327                                        \\
& Gene   Names      & \makecell[l]{HSD17B1, E17KSR, EDH17B1, EDH17B2,\\ EDHB17, SDR28C1} \\
& Protein Type                   & Enzymes                       \\
& Species                        & \textit{Homo sapiens} (Human)          \\
& EC Number                      & 1.1.1.62, 1.1.1.51            \\
& Hits for all PROSITE motifs  & PS00061                       \\
& Disprot ID                     & DP00023                       \\ 
\midrule
\multirow{15}{*}{MaSIF Profile}      
& Number of total surface vertex & 7265                           \\
& Number of Interacting face vertex  & 1061                     \\
& Interacting face region proportion & 0.146                   \\
& Number of hydrophilic vertex   & 933                            \\
& Hydrophilic region proportion  & 0.128                          \\
& Number of hydrophobic vertex   & 379                            \\
& Hydrophobic region proportion  & 0.052                          \\ 
& Number of Hbond doner vertex  & 232                         \\ 
& Hbond doner region proportion  & 0.032                          \\ 
& Number of Hbond receptor vertex  & 400                          \\ 
& Hbond receptor region proportion  & 0.055                          \\ 
& Number of positive charge vertex  & 400                          \\ 
& Positive charge region proportion  & 0.055                          \\ 
& Number of negative charge vertex  & 816                          \\ 
& Negative charge region proportion  & 0.112                          \\ 
\midrule
\multirow{8}{*}{NucleicNet Profile} 
& Number of Ribose                   & 882                            \\
& Number of Phosphate                    & 925                            \\
& Number of Guanine                   & 501                            \\
& Number of Uracil                    & 200                            \\
& Number of Adenine                   & 175                            \\
& Number of Cytosine                   & 188                            \\
\bottomrule
\end{tabular}
\label{table1}
\end{table*}

\begin{figure*}[p]
\includegraphics[width=0.85\textwidth]{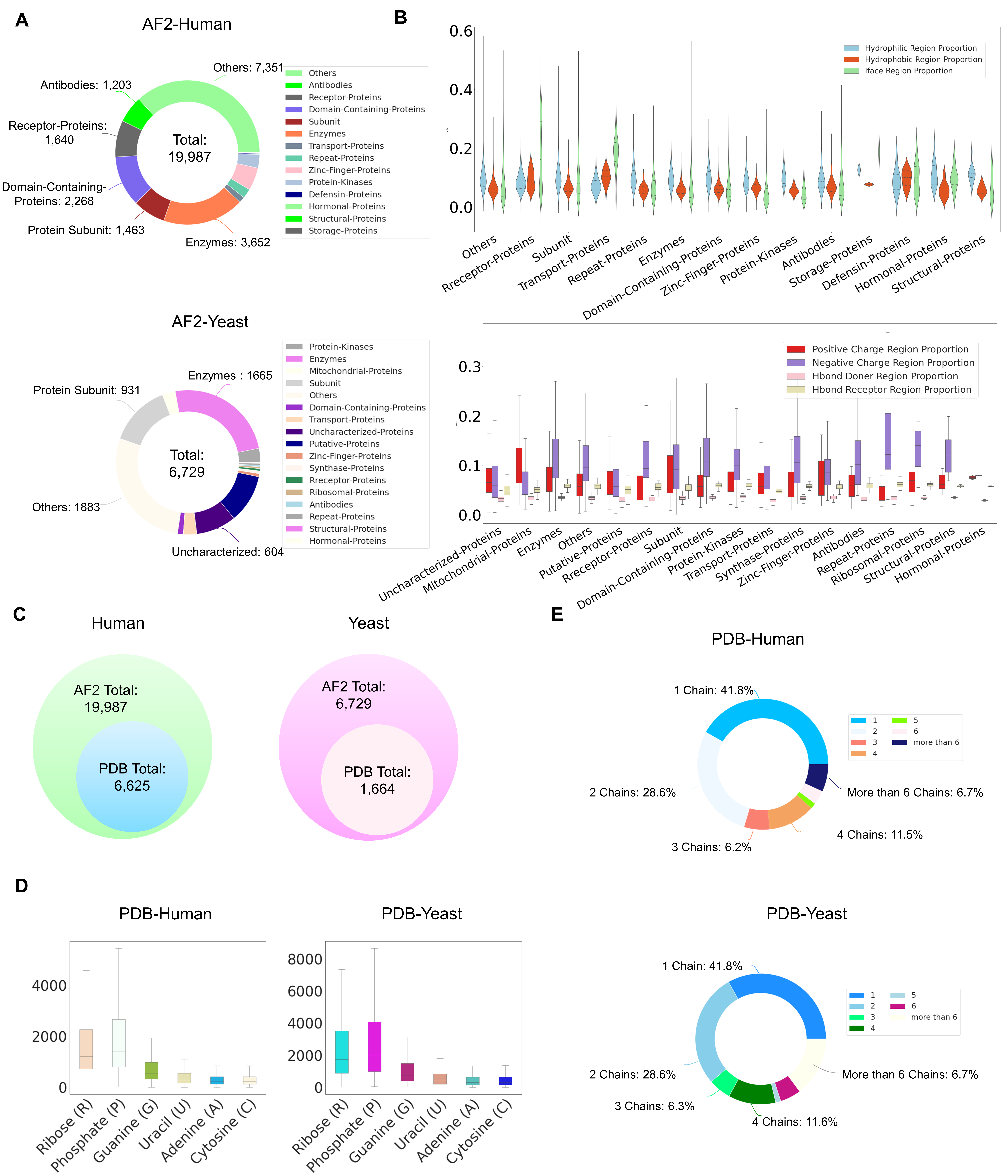}
\centering
\caption{
ProNet DB statistics for both Human and Yeast results in AlphaFold DB and PDB. (\textbf{A}) The functional classification for protein structures in both AlphaFold DB and PDB. (\textbf{B}) The upper panel illustrates the protein surface physicochemical property distribution including hydrophilic, hydrophobic, and interacting face region proportion of Human protein surface in AlphaFold DB. The beneath panel reveals the distribution of the positive/negative charge region and the Hbond Doner/Receptor region proportion of Yeast protein surface in AlphaFold DB. (\textbf{C}) Venn diagram shows the number of experimentally validated protein structures from PDB, compared with computationally predicted structures from AlphaFold DB. (\textbf{D}) Detailed comparison of the proportion of binding profiles of each RNA constituent in PDB, \textit{e.g.}, 4 bases: Adenine(A)/ Guanine(G)/ Cytosine(C)/ Uracil(U), and 2 backbone constituents: phosphate (P) and ribose (R). (\textbf{E}) The proportion of the number of chains in the PDB database in Human and Yeast.}
\label{statistic_fig}
\end{figure*}

\section*{Data Records}

All metadata and processed data are now available, and several approaches are provided to access the data. Totally, ProNet DB collected 16 model organism proteomes with 333,365 protein structures, covering various protein surface feature representation including g hydrophobicity, charge distribution, hydrogen bond, interacting face, and RNA-binding landscape such as RNA binding sites and RNA binding preference. A dedicated web page is for downloading archived file \textcolor{blue}{\href{https://proj.cse.cuhk.edu.hk/aihlab/pronet/\#/download}{https://proj.cse.cuhk.edu.hk/aihlab/pronet/\#/download}} and these data are also publicly available from Figshare repository \cite{pronetdataset}, including the original PDB format files from both PDB and AlphaFold DB, and protein surface representation result files in .PLY, .NPY and .PSE formats through MaSIF and NucleicNet. For each individual protein, a public API linked by UniProt ID is available and provides all meta information details and download links for a separate file. For example, for UniProt ID A2P2R3, its MaSIF information page is \textcolor{blue}{\href{https://proj.cse.cuhk.edu.hk/pronet/masifdata.html\#/entry/A2P2R3}{https://proj.cse.cuhk.edu.hk/aihlab/pronet/masifdata.html\#/entry/A2P2R3}}, and the NucleicNet information page is \textcolor{blue}{\href{https://proj.cse.cuhk.edu.hk/pronet/dataview.html\#/entry/A2P2R3}{https://proj.cse.cuhk.edu.hk/aihlab/pronet/dataview.html\#/entry/A2P2R3}}.

\section*{Technical Validation}

\subsection*{Database Statistics}
\begin{figure*}[ht]
\includegraphics[width=0.85\textwidth]{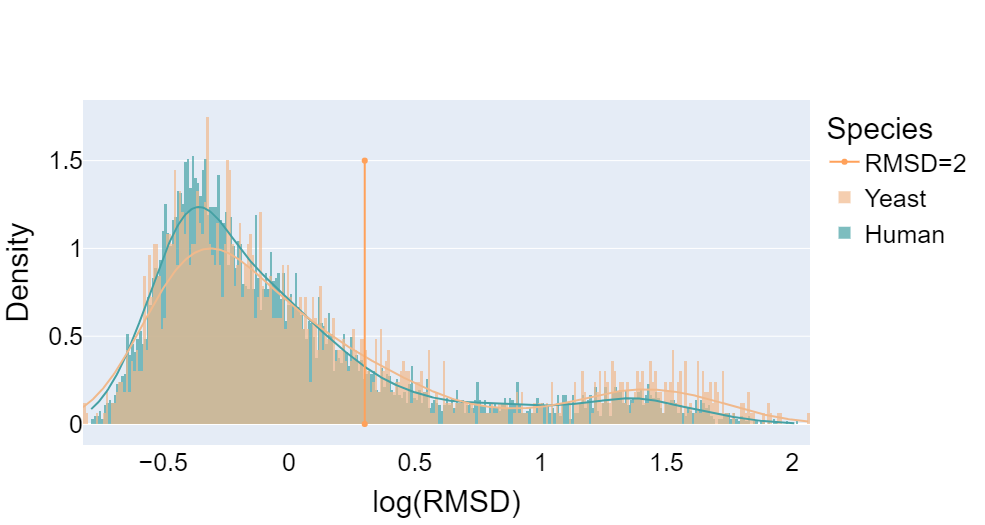}
\centering
\caption{ RMSD statistics with $log2$ as the threshold. The majority of the RMSD values calculated from protein structures in PDB Human(5,325/6,600) and Yeast(1,242/1,660) falls under the threshold $log2 \approx 0.30102$). 
}
\label{supp_fig_rmsd}
\end{figure*}

Currently, the database contains over 16 model organism species and 333,365 entries for proteome (Table \ref{proteomes}) from AlphaFold DB and PDB (\textit{Homo sapiens}: 23,391 and 6,030; \textit{Saccharomyces cerevisiae}: 6,040 and 1,160, and other 14 model organism species.). We further classify these proteins by their function into sub-classes, such as antibodies and enzymes. For \textit{Homo sapiens}, the protein structures are classified into 15 sub-classes: Antibodies, Contractile-Proteins, Enzymes, Hormonal-Proteins, Structural-Proteins, Storage-Proteins, Transport-Proteins, Zinc-Finger Proteins, Receptor Proteins, Domain-Containing-Proteins, Defensin-Proteins, Repeat-Proteins, Subunit, and Protein-Kinases. Those proteins with unknown categories are classified as \textit{Others}. From Figure \ref{statistic_fig} (A), although the majority of the structures are labeled as unknown (7,351 entries), we see that the majority of the human protein types are clustered in Enzymes (3,652 entries), Domain-Containing-Protein (2,268 entries), and Preceptor-Proteins (1,640 entries). Figure \ref{statistic_fig} (C) shows that a certain number of protein structures (66.9\% in \textit{Homo sapiens}, 75.2\% in \textit{Saccharomyces cerevisiae}) have not been validated by experimental approaches. Besides, the Supplementary Figure \ref{supp_fig_rmsd} demonstrates the considerable protein structure prediction performance, since 80.6\% of validated proteins in \textit{Homo sapiens} and 74.8\% of validated proteins in \textit{Saccharomyces cerevisiae} are accurately predicted (RMSD $\leq$ 2.0). In Figure \ref{statistic_fig} (B), we integrate the proportion for hydrophobic and hydrophilic vertex over the total number of vertex for AlphaFold2 Human proteins and compare them with the interacting face proportion. A clear pattern is shown in Figure \ref{statistic_fig} (B) that the Hbond receptor region is statistically higher than Hbond donor region in AlphaFold Yeast proteins. In addition, Figure \ref{statistic_fig} (D) reveals that the overall number of nucleic acids for Yeast is much higher than that of human. The statistical results of the chain numbers from both PDB Human and Yeast data are illustrated in Figure \ref{statistic_fig} (E). More statistic result of Human and Yeast data can be found in Supplementary Figure \ref{supp_fig_stats} and the other model organism proteomes information is in Supplementary Table \ref{Alphafold2_species_statistics}. 

\subsection*{Case Study}

\begin{figure*}[t]
\includegraphics[width=0.9\textwidth]{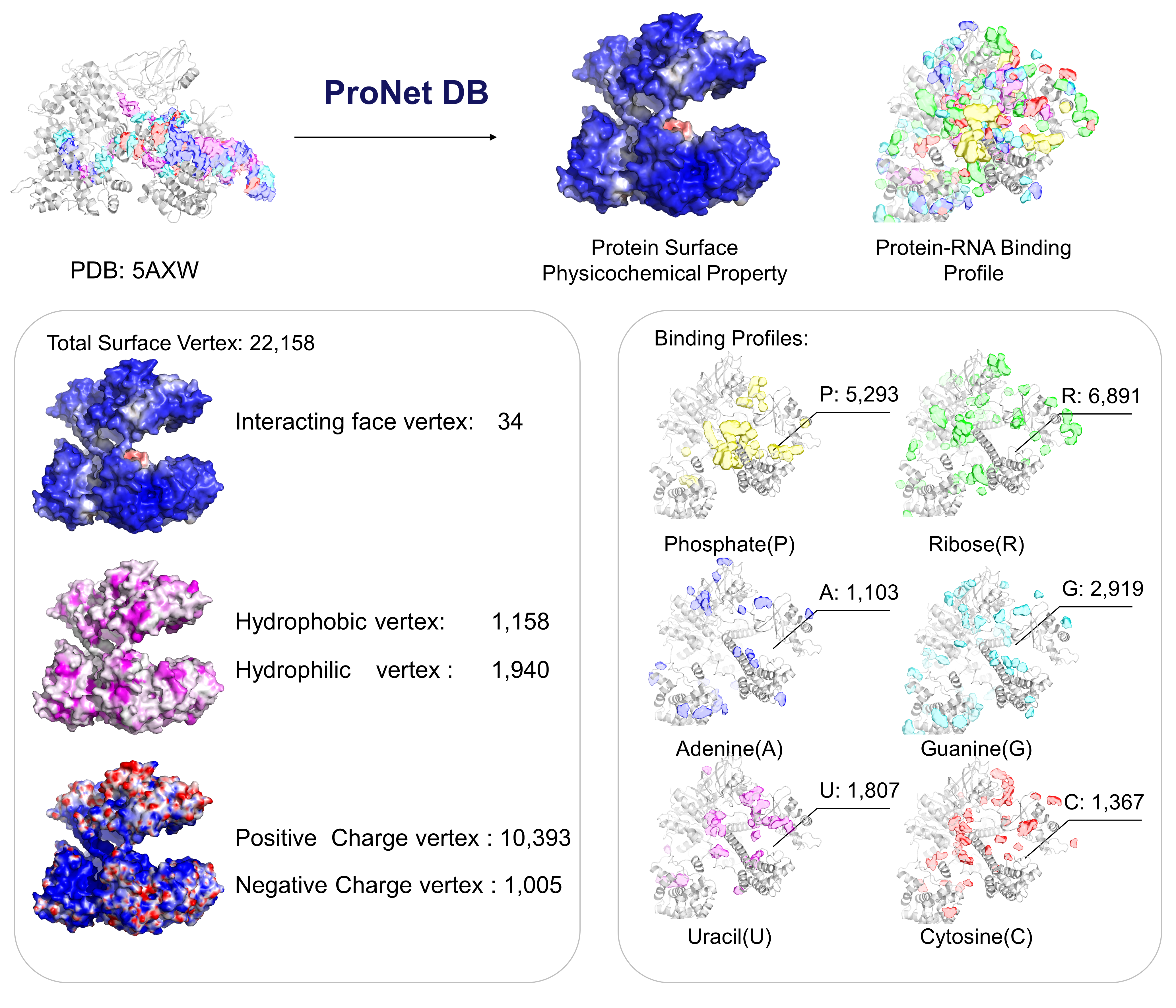}
\centering
\caption{Case study (PDB: 5AXW): ProNet DB shows comprehensive information of the protein structure surface fingerprints as well as the protein RNA binding landscape. On the left panel: Iface region is consistent with the nucleic acid binding sites, and electron donor region is located at non-binding sites. On the right panel: The RNA binding landscape shows the RNA binding sites located at the inner region.}
\label{case_study}
\end{figure*}

Here, we employed CRISPR/Cas gene editing system for case study since it is a well-established protein-nucleic acid interaction issue and a variety of structural analysis studies focus on Cas9 protein. Cas9 protein plays an important role in the CRISPR-Cas system, and thus understanding how Cas9 mediates RNA-guided DNA recognition is the essential part to improve the gene editing system. The crystal structure of \textit{Staphylococcus aureus} Cas9 (PDB: 5AXW) was chosen for protein surface physicochemical property and RNA-binding profiles analysis.

As shown in Figure \ref{case_study}, we have highlighted the SaCas9–sgRNA chimeric complex structure with its binding guide RNA, in which a central channel was formed in the middle of the structure. From the protein surface fingerprint results of Iface region, ProNet DB reveals that the original nucleic acid binding site is located at the interacting face region, compared to the non-binding site. Besides, the electron donor region shows the inner region is positive-charged, indicating the interaction between protein surface and nucleic acids in the central channel between the recognition and nuclease lobes \cite{jinek2014structures,nishimasu2014crystal}. In Figure  \ref{case_study}, the predicted protein-RNA binding profiles with specific RNA binding site and binding preference indicate that the corresponding RNA molecule is located in the inner region of protein structure, which is consistent with the ground truth. These results show that such \textit{in silico} approach can efficiently capture the protein physicochemical surface property as well as RNA binding landscape and lay the foundation for downstream tasks such as sgRNA design \cite{doench2016optimized} and CRISPR system optimization \cite{hu2021discovery}.

\section*{Usage Notes}

Multi-scale data analyses for 333,365 protein structures and 16 model organism species were processed under diverse protein functional categories. The homepage (Figure \ref{overview_withlabel}) integrates all the server tools and is divided into three major components: prediction tools, database queries, and visualization tools. An overview and interactive table present information ranging from protein name, PDB ID, UniProt ID, protein type, interacting face proportion, Hbond region proportion, positive/negative charge region proportion, and protein-RNA-binding profiles (see Table \ref{table1}). The web interface was established to search for the property profiles of a user-specified protein type in multiple species and the search browser is shown in the top-right part of Figure \ref{overview_withlabel}. The protein information page provides detailed information, download link of processed protein surface feature and visualization and more information can be found on \textcolor{blue}{\href{https://proj.cse.cuhk.edu.hk/aihlab/pronet/\#/services}{https://proj.cse.cuhk.edu.hk/aihlab/pronet/\#/services}}.

\section*{Code availability}

The process code is available at \textcolor{blue}{\href{https://github.com/jxmelody/PronetProcess}{https://github.com/jxmelody/PronetProcess}}. 
\\The ProNet DB link is \textcolor{blue}{\href{https://proj.cse.cuhk.edu.hk/aihlab/pronet/\#/Home}{https://proj.cse.cuhk.edu.hk/aihlab/pronet/\#/Home}}.\\
All primary data is uploaded to Figshare \textcolor{blue}{\href{https://figshare.com/s/83bc43fac5aec6d1e0e6}{https://figshare.com/s/83bc43fac5aec6d1e0e6}}.



\section*{Competing interests}

None declared

\section*{Author Contributions}

J.W. conducted structure collection and MaSIF analysis; J.X. set up web server and structure visualization; S.C. implemented NucleicNet analysis and statistical analysis; L.Z. performed protein information collection. X.G. and Y.L conceived the study; All authors wrote the manuscript. 

\beginsupplement
\begin{figure*}[ht]
\includegraphics[width=0.85\textwidth]{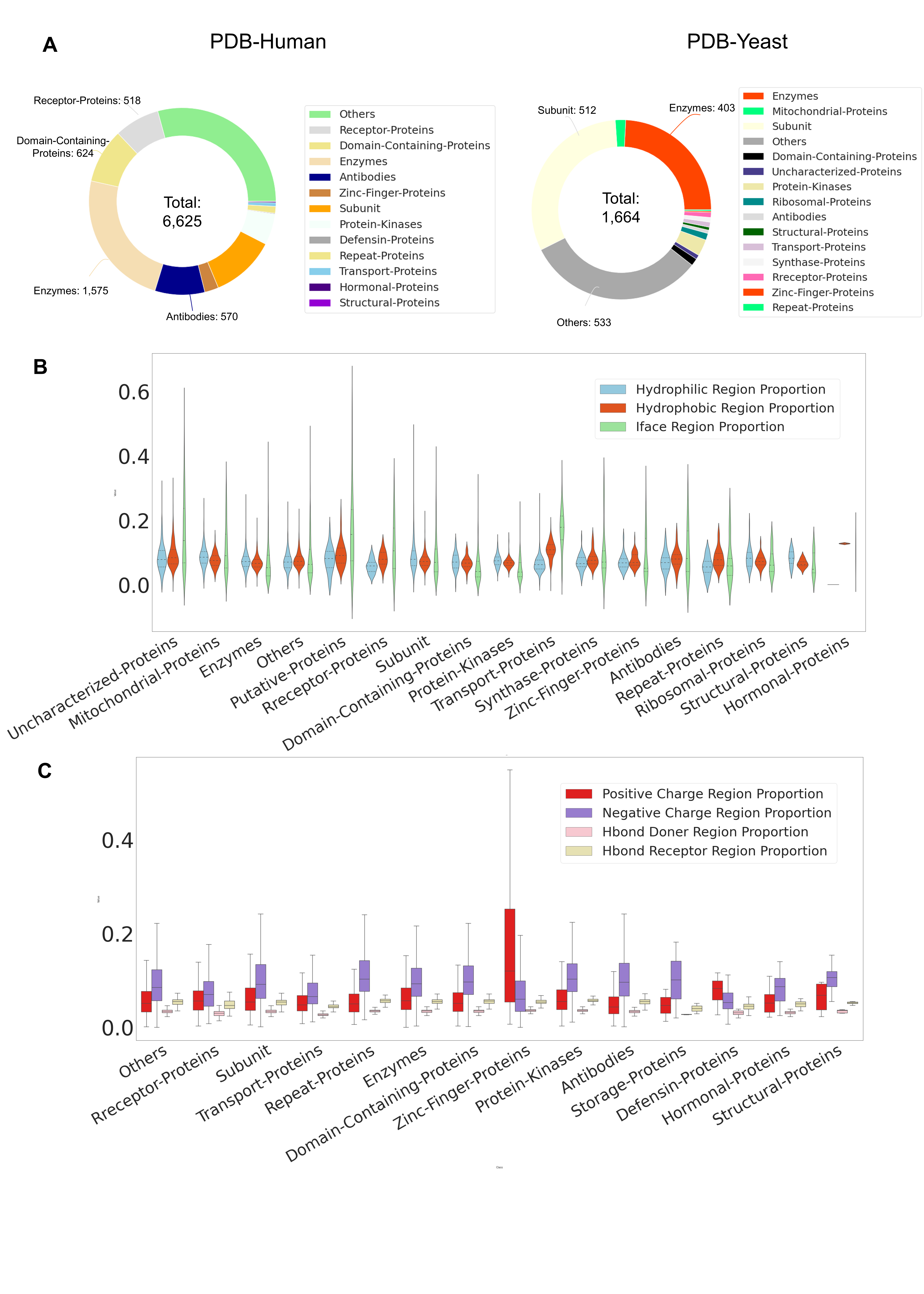}
\centering
\caption{ Supplementary statistical results for ProNet DB. (\textbf{A}) The functional classification for protein structures in PDB Human and Yeast. (\textbf{B}) Protein surface physicochemical property distribution including hydrophilic, hydrophobic, and interacting face region proportion of Human protein surface in PDB. (\textbf{C}) The distribution of the positive/negative charge region and the Hbond Doner/Receptor region proportion of Yeast protein surface in PDB.
}
\label{supp_fig_stats}
\end{figure*}

\begin{sidewaystable*}[htb]
    \caption{Detailed data statistics for ProNetDB}
    \label{Alphafold2_species_statistics}
    \renewcommand{\arraystretch}{3}
    \centering
    \resizebox{\textwidth}{!}{%
    \begin{tabular}{|c|cccccccccc|}
\hline
Species & \multicolumn{10}{c|}{\textbf{Protein Types (Top 10 categories)} }  \\

\hline
{\multirow{2}*{\makecell{Candida albicans \\ 5,974}}} & \makecell{Domain-Containing \\Proteins} & Enzymes & Subunit &  \makecell{Protein\\Kinases} & \makecell{Mitochondrial\\Proteins} & \makecell{Ribosomal\\Proteins}&\makecell{Transport\\Proteins} & \makecell{Putative\\Proteins} & \makecell{Membrane\\Proteins} & \makecell{Others}\\
\cline{2-11}
~ & 1,093 (18.3\%) & 961(16.08\%) & 612(10.3\%) & 164(2.7\%) & 159(2.6\%) & 99(1.7\%) & 94(1.6\%)& 71(1.2\%)  & 60(1.0\%) & 2,304(38.6\%) \\
\hline

{\multirow{2}*{\makecell{Arabidopsis thaliana \\ 27,427}}} & Enzymes &   \makecell{Domain-Containing \\Proteins}&  \makecell{Repeat \\Proteins} &  \makecell{Subunit} & \makecell{Putative\\Proteins} & \makecell{Protein\\Kinases}&\makecell{Zinc-Finger\\Proteins} & \makecell{Rreceptor\\Proteins} & \makecell{Membrane\\Proteins} & \makecell{Others}\\
\cline{2-11}
~ & 5,858 (21.3\%) & 2,348(8.56\%) & 1,272(4.6\%) & 1,231(4.5\%) & 921(3.4\%) & 841(3.06\%) & 724(2.6\%)& 699(2.5\%)  & 675(2.4\%) & 8,077(29.6\%) \\
\hline

{\multirow{2}*{\makecell{Escherichia coli \\ 4,289}}} & Enzymes &   \makecell{Subunit}&  \makecell{Transport \\Proteins} &  \makecell{Membrane\\Proteins} & \makecell{Putative\\Proteins} & \makecell{Protein\\Kinases}&\makecell{Antibodies} & \makecell{Synthase\\Proteins} & \makecell{Zinc-Finger\\Proteins} & \makecell{Others}\\
\cline{2-11}
~ & 1,528 (35.2\%) & 337(7.8\%) & 247(5.7\%) & 164(3.8\%) & 129(3\%) & 117(2.7\%) & 65(1.5\%)& 37(0.8\%)  & 21(0.4\%) & 1,019(23.7\%) \\
\hline

{\multirow{2}*{\makecell{Danio rerio \\ 24,664}}} & Enzymes &   \makecell{Domain-Containing \\Proteins}&  \makecell{Rreceptor \\Proteins} &  \makecell{Subunit} &\makecell{Protein\\Kinases}& \makecell{Zinc-Finger\\Proteins} & \makecell{Membrane\\Proteins} & \makecell{Repeat\\Proteins} & \makecell{Transport\\Proteins} & \makecell{Others}\\
\cline{2-11}
~ & 3,348 (13.57\%) & 2,728(11.1\%) & 1,673(6.7\%) & 1,154(4.6\%) & 881(3.57\%) & 746(3.0\%) & 420(1.7\%)& 371(0.8\%)  & 289(1.1\%) & 12,128(49.1\%) \\
\hline

\end{tabular}}
\end{sidewaystable*}

\bibliography{sample}

\end{document}